\documentclass[twocolumn,secnumarabic,amssymb,amsmath, nobibnotes, aps, prb]{revtex4-1}

\usepackage{tabularx}
\usepackage{float}
\usepackage{bm}

\usepackage{graphicx}% Include figure files
\usepackage{amsmath}
\usepackage{color}
\usepackage{hyperref}
\hypersetup{
    colorlinks=false,       % false: boxed links; true: colored links
    linkcolor=blue,          % color of internal links (change box color with linkbordercolor)
    citecolor=red,        % color of links to bibliography
    filecolor=magenta,      % color of file links
    urlcolor=red           % color of external links
}

\begin{document}

\title{Core-shell multilayered nanoparticles: giant photonic density of states coupled to the far-field.}

\author{Mehedi Hasan$^1$}
\author{Ivan V. Iorsh$^1$}%
\email{i.iorsh@phoi.ifmo.ru}
\author{Pavel A. Belov$^1$}
\affiliation{
$^1$The Metamaterials Laboratory, National Research University of Information Technologies, Mechanics and Optics (ITMO), St. Petersburg 197101, Russia
}

\begin{abstract}
We present a quantum theoretical treatment of light-matter coupling in the system consisting of a quantum dot and a spherical core-shell metal-dielectric multilayer nanoparticle. It is shown that both weak and strong coupling regimes can be realized in the set-up. Specifically, we demonstrate a strong coupling regime between a quantum dot and a nanoparticle, when the quantum dot resonance is tuned to the frequency at which normal component of effective nanoparticle permittivity is crossing zero. Moreover, we demonstrate the regime at which the quantum dot decays   much faster than in vacuum (due to the large Purcell factor) and at the same time radiates  more power to the far field. This findings pave the way towards more efficient control over radiation properties of quantum emitters.

\end{abstract}

\maketitle
%\section{Introduction}
Tailoring the radiation properties of the quantum emitters by placing them inside or in the vicinity of the microcavities has recently emerged in a rapidly evolving field known as Cavity Quantum Electrodynamics~\cite{kavokin,deliberato,beaudoin,vidal-prl,vvlac,delga,wang,shalabny}. Typical solid state realizations of  microcavities are based on photonic crystals with defect(s),\cite{akahane} or Bragg mirrors~\cite{reithmaier}. To boost the light-matter interaction in the sub-wavelength regime, there has been some on-going interest in the realization of strong light-matter coupling with plasmonic cavities that are well-known for confining light below the diffraction limit(see Ref.~\cite{torma}, for a comprehensive review).

Depending on the degree of coupling between the emitter and the external medium, the situation can be classified into `weak-coupling' and `strong-coupling'. In the former case, energy propagation from a quantum emitter to it's environment is unidirectional, i.e. energy of the emitter decays in a monotonous way, while in the latter case, the emitted energy is reflected back to the original emitter and then re-emitted by the emitter and so on, thus the energy transfer is bi-directional. The QE can be a quantum dot (QD), fluorescent molecule, or  a single atom. In the study of weak-coupling, the external body near the emitter is considered as perturbation to the original system -- atom-in-vacuum. However, to understand the full physics in the strong-coupling regime, non-perturbative calculations are imperative. To date, strong-and ultra-strong-coupling have been studied for wide variety of systems~\cite{savasta, zubairy, chantharasupawong, wu-oi, cacciola, antosiewicz, uemoto, thongrattanasiri}. In this Communication, we consider the coupling of a quantum emitter to a multilayered spherical core-shell nanoparticles.

Core-shell nanoparticles~\cite{GhoshChaudhuri2011} have recently become an extremely popular object of study in nanophotonics. Namely, core-shell nanoparticles have been suggested for the realization of both super-scattering and total scattering cancellation~\cite{Chen2012}. Moreover, it has been shown that the multilayered metal-dielectric core-shell nanoparticles, can be used for the realization of spherical hyperbolic cavities, i.e. cavities made from hyperbolic metamaterials~\cite{Rho2010}.
Hyperbolic metamaterials are a particular class of metamaterials where the dielectric tensor has indefinite signature, and the iso-frequency surface has infinite volume that leads to broadband singularity of photonic local density of states (LDOS), which results in the broadband increase of the Purcell factor for the emitters planced in the vicinity of hyperbolic metamaterials~\cite{cortes,zubin,vanya,puddubny_pra,puddubny_springer,krishnamoorthy}. 
At the same time for the case of hyperbolic metamaterials, the enhanced Purcell factor does not lead to the enhancement of the radiation intensity of the emitters placed in its vicinity. Contrary to that, radiation intensity is actually decreased, since most of the emitter energy is transferred to the near-field modes of the metamaterial, which decay non-radiatively. In this Communication we show that substitution of  the bulk hyperbolic metamaterial with a finite-size hyperbolic cavity can provide both large Purcell factor and radiation intensity increase for the quantum emitters placed in its vicinity.
\begin{figure}[!h]
\centering
\includegraphics[width=0.45 \textwidth]{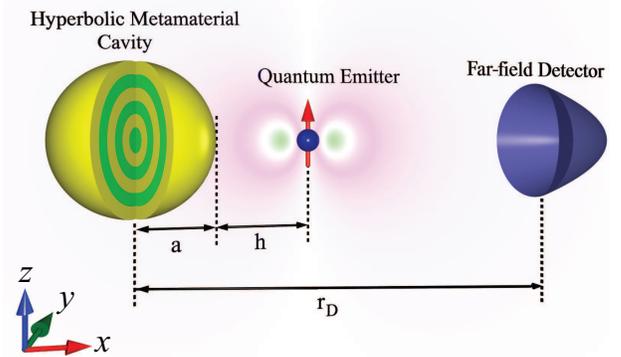}
\caption{A hyperbolic cavity of spherical shape (radii $a$) is in the vicinity of a quantum dot at a distance $h$ from the surface of the sphere. The far-field spectrum is attained via a detector at distance $r_\mathrm{D}$ from the center of the sphere. The above schematic shows a `cut-view' of the sphere.}
\label{fig:fig_1}
\end{figure}

While a considerable number of papers have been dedicated to the studies of the Purcell effect in hyperbolic metamaterials, the model of the strong coupling of the quantum emitter to the hyperbolic cavity has not been put forward so far as to our knowledge. 

In this Communication, we report that a hyperbolic cavity and a quantum emitter exchange photon in the strong-coupling regime. Via non-perturbative calculation, we reveal the existence of two distinct anti-crossings in the local polarization spectrum: one at the vicinity of the nanoparticle pseudomode frequency formed by the series of high angular momentum modes, and one in the vicinity of the frequency, at which the metal-dielectric cavity experiences the topological transition -- from the elliptical to hyperbolic regime. However only in the latter case, the radiation reaches the far field zone. Moreover, the mode that is detectable at far-field has quite large Purcell factor. We report an unique behavior of the strong-coupled system -- radiation power enhancement and large Purcell factor at the same spectral position in hyperbolic metamaterial cavity. 

%\textbf{This paper is organized as follows: in the next section, the theoretical model for the treatment of strong-coupled system is outlined along with the perturbative Purcell factor and Lamb shift; then the non-perturbative polarization spectrum and far-field spectrum is presented. Finally, in the  third section some conclusions are drawn based on the results in the aforementioned sections.}

%Should we remove this?

\begin{figure}[b]
\centering
\includegraphics[width=0.5 \textwidth]{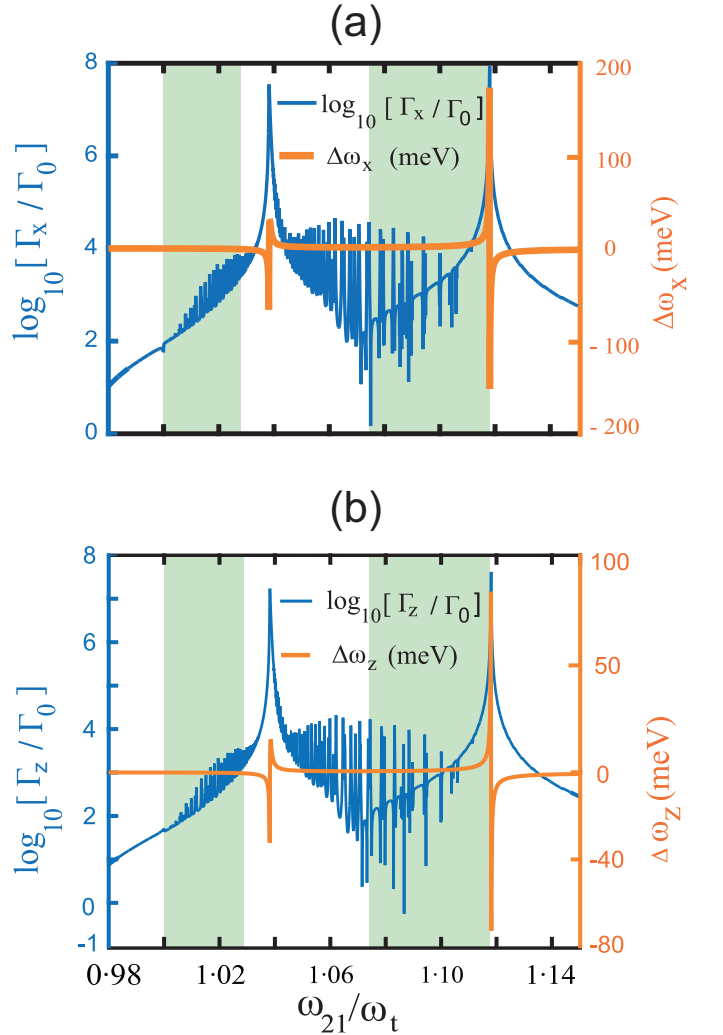}\\
\caption{Purcell factor and Lamb shift for a dipole along (a) $x$-directed dipole, (b) $z$-directed dipole, for a hyperbolic cavity shown in Fig.~\ref{fig:fig_1}. Since, the quantum dot has been placed at a very close proximity to the sphere, a  number of modes is excited by the QD that give rise to a series of  narrow peaks along with the peak at epsilon-near-zero frequency ($\omega_{21}/\omega_t = 1.118$) and at $\omega_{21}/\omega_t = 1.038$. The shaded green regions represent the hyperbolic regime. Note the logarithmic scale of the left axis for Purcell factor.}
\label{fig:fig_2}
\end{figure}

%\section{Model and Results} \label{sec:theory}
In order to understand the radiation properties of the quantum emitter to the environment around it as shown in Fig~\ref{fig:fig_1}, the Purcell factor $\Gamma/\Gamma_0$ and Lamb shift $\Delta\omega$ are calculated upto the leading order of perturbation which are shown in Fig.~\ref{fig:fig_2}, for two different orientations of quantum dot. The quantum emitter is a point quantum dot and is placed  at a distance $h$ from the surface of the spherical cavity made of alternating dielectric and metal concentric layers of fixed filling fraction and the following parameters are fixed throughout this paper unless otherwise mentioned: the sphere consists of ten periods  and the permittivity of the dielectric is $\varepsilon_1 = 2.25$ and the metal is modeled by Lorentz model, $\varepsilon(\omega)=1+\frac{\omega_p^2}{\omega_t^2-\omega^2-i\omega\gamma}$; the parameters for the model are -- $\omega_p/\omega_\mathrm{t}=0.5$ and $\gamma/\omega_\mathrm{t}=10^{-4}$ and $h = 10^{-3}\times \lambda_{\mathrm{t}}$, where $\lambda_{\mathrm{t}}=2\pi c/\omega_{\mathrm{t}}$. The radius of the multilayered particle is $0.1\times \lambda_\mathrm{t}$, the filling fraction of metal in one period is $0.5$, and the linewidth of the quantum dot is $10^{-4}\times\omega_t$. 

Purcell factor can be calculated via
\begin{equation}
\label{eq:purcell}
\Gamma = \Gamma_0 + \frac{2\check{\omega}_{21}^2}{\hslash\varepsilon_0 c^2} \Im\left[\textbf{d}\cdot\widehat{\textbf{G}}(\textbf{r}_\mathrm{QD},\textbf{r}_\mathrm{QD},\omega_{21})\cdot\textbf{d}\right],
\end{equation}
where $\Gamma_0$ is spontaneous decay rate in vacuum, $\check{\omega }_{21}$ is quantum dot frequency, $\omega_{21}$ plus the Lamb shift, $\Delta\omega$. In the leading order of perturbation, $\check{\omega}_{21}$ can be replaced by `bare QD frequency', $\omega_{21}$ and $\textbf{d}$ is dipole matrix element of QD while $\widehat{\textbf{G}}(\textbf{r}, \textbf{r}_\mathrm{QD}; \omega)$ is the Green's function of the system -- 
\begin{equation}
\label{eq:green's_curl}
\left[\nabla\times\nabla\times - \frac{\omega^2}{c^2}\widehat{\varepsilon}(\omega, \textbf{r})\right]\widehat{\textbf{G}}(\textbf{r}, \textbf{r}_\mathrm{QD}; \omega) = \mathrm{\widehat{I}}\delta(\textbf{r} - \textbf{r}_\mathrm{QD}),
\end{equation}
$\widehat{I}$ being the usual $3\times3$ unit dyad. The Green's function in Eq.~\ref{eq:green's_curl} is calculated by expansion of the field as a series of spherical harmonics~\cite{Li}.

The Lamb shift reads as  
\begin{equation}
\label{eq:lamb}
\Delta\omega = -\frac{\Re\left[\textbf{d}\cdot \widehat{\textbf{G}}(\textbf{r}_\mathrm{QD}, \textbf{r}_\mathrm{QD}; \omega) \cdot \textbf{d}\right]}{\hslash \varepsilon_0}\cdot\frac{\omega^2_{21}}{c^2}.
\end{equation}

Since the quantum dot has been placed at a very close proximity to the spherical cavity (compared to the wavelengths of interest), a number of modes is excited by the QD that give rise to a series of narrow  peaks in addition to the primary peaks at $\omega_{21}/\omega_t \approx 1.04 \mathrm{and} 1.12$. The second dominant peak ($\omega_{21}/\omega_\mathrm{t}\approx 1.12$) in the Purcell factor and Lamb shift can be understood via the effective medium theory for metamaterial homogenization that reads --
\begin{equation}
\begin{array}{l}
{\varepsilon _{\theta ,{\rm{ }}\varphi }} = {d_1}{\varepsilon _m} + {d_2}{\varepsilon _d};\\
\varepsilon _r^{ - 1} = {d_1}\varepsilon _m^{ - 1} + {d_2}\varepsilon _d^{ - 1}.
\end{array}
\end{equation} 
where $\varepsilon_r$ is the radial component of the dielectric tensor and $\varepsilon_{\theta,\varphi}$ the transverse component of the dielectric tensor. Two regions can be distinguished from the above description, namely 1) elliptical regime where $\varepsilon_r$ and $\varepsilon_{\theta,\varphi}$ have same sign, 2) hyperbolic regime where $\varepsilon_r$ and $\varepsilon_{\theta,\varphi}$ have opposite signs. The hyperbolic regime is shaded by green in Fig.~\ref{fig:fig_2}, the unshaded portion represents the elliptical regime. The enhanced Purcell factor around $\omega/\omega)_t\approx1.12$ occur at the the spectral point where hyperbolic to elliptical topological transition occurs. As will be discussed later that the first resonant peak originates from the  hybridization of higher order spherical harmonics. The Lamb shift for the $x$ and $z$-oriented dipole momentum is rather large compared to typical Lamb shift in experiment. \cite{noginov} This high Purcell factor and large Lamb shift is the initial signature of strong-coupling that requires further scrutiny via non-perturbative calculation. 

Before embarking on the non-perturbative calculation, we would like to mention one little subtlety with the real part of the Green's function due to it's diverging nature. However, this divergence can be gotten rid of by setting high-$k$ cut-off \cite{novotny}. This regularization extracts realistic Lamb shift because any finite-sized dipole must have a finite Lamb shift~\cite{bethe}. Moreover, near the resonance, the homogeneous vacuum contribution is orders of magnitude smaller than the scattering part of the Green's function. Thus disregarding the homogeneous part of Green's function is a reasonable approximation and this allows one to get rid of the unphysical divergence of Lamb shift. In fact, if the whole calculation is performed using Lippmann–-Schwinger approach, the divergent part of Green's function is actually gotten rid of and it is the scattering part that contributes eventually~\cite{vidal-prl,vvlac}. Hence, instead of total Green's function $\widehat{\textbf{G}}(\textbf{r}, \textbf{r}; \omega)$ that consists of vacuum Green's finction $\widehat{\textbf{G}}^0(\textbf{r}, \textbf{r}; \omega)$ and the scattering part of the Green's function $\widehat{\textbf{G}}^\mathrm{sc}(\textbf{r}, \textbf{r}; \omega)$, it is reasonable to replace $\widehat{\textbf{G}}(\textbf{r}, \textbf{r}; \omega)$ by $\widehat{\textbf{G}}^\mathrm{sc}(\textbf{r}, \textbf{r}; \omega)$ in Eq.~\ref{eq:purcell} and \ref{eq:lamb}.

Now, to study the coupling properties in a non-perturbative fashion, let us consider a state-space $\mathcal{H}$ with two states \cite{welsh,vidal-prl} namely, the ground state, $\vert1\rangle = \left(
\begin{array}{c}
 0 \\
 1 \\
\end{array}
\right),$ and the excited state, $\vert2\rangle = \left(
\begin{array}{c}
 1 \\
 0 \\
\end{array}
\right).$ 

Considering weak excitation, i.e. only one quantum of light is associated at a time, the Hamiltonian of the system can be written as $-$
\begin{equation*}
\begin{split}
\widehat{\mathcal{H}} = \sum\limits_{\lambda = e, m} \int \! \mathrm{d}\textbf{r} \int_0^\infty \! \mathrm{d}\omega_l~\hslash \omega_l~\widehat{\textbf{f}}^\dagger_\lambda(\textbf{r}, \omega_l) ~ \widehat{\textbf{f}}_\lambda(\textbf{r}, \omega_l)\\
-\left[ \widehat{\sigma}^+ \textbf{d} + \widehat{\sigma}^- \textbf{d}\right]\cdot\widehat{\textbf{F}}(\textbf{r}_\mathrm{QD}) + \hslash \omega_{21} \widehat{\sigma}^+\widehat{\sigma}^-,
\end{split}
\end{equation*}
where $e$ and $m$ stand for electric and magnetic field, respectively, that are related to the continuum Bosonic-field operator $\widehat{\textbf{f}}(\textbf{r}, \omega_l)$ and it's Hermitian conjugate $\widehat{\textbf{f}}^\dagger(\textbf{r}, \omega_l)$ and, with eigenfrequency $\omega_l$. The operators $\widehat{\sigma}^+ = \vert2\rangle\langle1\vert$ and $\widehat{\sigma}^- = \vert1\rangle\langle2\vert$ are the usual Pauli operators for exciton, and $\textbf{d}$ being the dipole matrix element (i.e., $\textbf{d} = \langle2|e\textbf{r}|1\rangle$) between the ground state, $|1\rangle$ and excited state, $|2\rangle$ with transition frequency, $\omega_{21}$.
Since, the primary interest of this work is in the optical regime, we set the relative permeability, $\mu = 1$ and exploiting bosonic operators, the field operator takes the following form -- 
\begin{equation}
\label{eq:fieldop}
\begin{split}
\widehat{\textbf{F}}(\textbf{r}) = \widehat{\textbf{E}}^0(\textbf{r}) + \frac{1}{\varepsilon_0}\cdot\frac{\omega^2}{c^2} \widehat{\textbf{G}}(\textbf{r}, \textbf{r}_\mathrm{QD}; \omega)\cdot \textbf{d} \\
\times\left[ \widehat{\sigma}^+(\omega)  + \widehat{\sigma}^-(\omega) \right],
\end{split}
\end{equation}
where $\widehat{\textbf{E}}^0(\textbf{r})$ is free-field operator, i.e. the field without the presence of quantum emitter. The local polarization spectrum that provides insight on quenching and other propagation mechanisms at the dipole position $\textbf{r}_\mathrm{QD}$ can be expressed via --
\begin{equation}
\label{eq:local-polarization}
\begin{aligned}
P(&\textbf{r}_\mathrm{QD}, \omega) \equiv \langle\widehat{\sigma}^+(\omega)\widehat{\sigma}^-(\omega)\rangle\\
            &= \bigg|\frac{1}{\omega_\mathrm{QD}^2 - \omega^2 - \frac{\omega^2}{c^2}\textbf{d}\cdot\widehat{\textbf{G}}(\textbf{r}, \textbf{r}_\mathrm{QD}; \omega)\cdot \textbf{d}/\hslash\varepsilon_0}\bigg|^2.\\
\end{aligned}
\end{equation} 

\begin{figure*}[!htb]
\centering
\includegraphics[width=0.90 \textwidth]{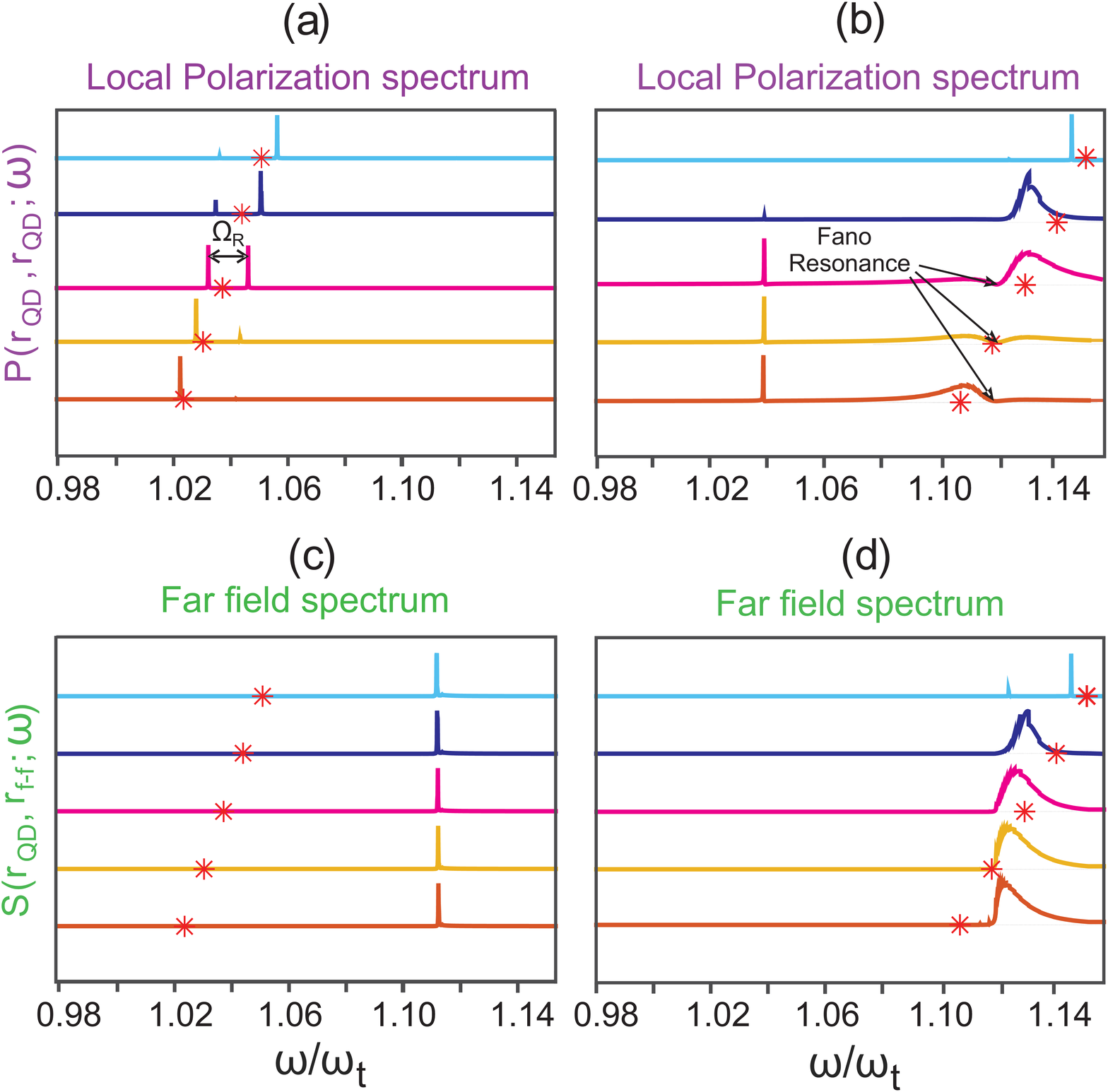}
\caption{(a) and (b), Local polarization spectrum of a quantum dot for different spectral position of quantum dot. (c) and (d) The far-field spectrum for the same spectral position of quantum dot same as (a) and (b). As shown here, the modes at $\omega/\omega_t\approx1.04$ does not appear at far-field while the mode near the topological transition of hyperbolic metamaterial. }
\label{fig:fig_3}
\end{figure*}

In Fig.~\ref{fig:fig_3}(a) and (b), the local polarization spectrum of the quantum dot are shown as a function of transition frequency of quantum dot, $\omega_{\mathrm{QD}}$. The `red-stars' represent spectral position of the quantum dot. As seen from Fig.~\ref{fig:fig_3} (a) and (b), two distinct anticrossings are evident, at two different spectral regimes. Note that the anticrossings in Fig.~\ref{fig:fig_3}(a) and (b) are centered around the two spectral points with high Purcell factor. This is the direct signature of strong-coupling of photon and HM. Despite the strong-coupling (i.e. strong light-matter energy exchange), the resonance at topological transition ($\omega/\omega_\mathrm{t} \approx 1.12$) is rather broad. This fact seems counter-intuitive. The resonance at the second anti-crossing has Fano line-shape: a slow varying broadband resonance interact with a narrow resonance that results the broad asymmetric line-shape. This explains the broad nature of the line-shape despite strong-coupling. 

To understand the distinctive nature of the two anticrossings the radiative property is further explored. The non-local spectrum of the QD, i.e., spectral density at a point $\textbf{r} (\neq\textbf{r}_\mathrm{QD})$ for a QD position at $\textbf{r}_\mathrm{QD}$ reads \cite{vvlac}, 
\begin{equation}
\label{eq:ff-spectrum-int}
\begin{split}
S(\textbf{r}, \omega) = \int_0^\infty\mathrm{d}t_1\int_0^\infty\mathrm{d}t_2\langle\mathrm{\widehat{F}}^\dagger(\textbf{r}, t_1)\mathrm{\widehat{F}}(\textbf{r}, t_2)\rangle\\
\cdot\mathrm{e}^{i\omega (t_2-t_1)}.\hspace{0.7in}
\end{split}
\end{equation}
Exploiting the field operator $\mathrm{\widehat{F}}(\textbf{r}, t_1)$ of Eq.~\ref{eq:fieldop}, the far-field spectrum in Eq.~\ref{eq:ff-spectrum-int} reads, 
\begin{equation}
\label{eq:ffspectrum}
S(\textbf{r},\textbf{r}_{QD}, \omega) = \bigg|\frac{\frac{\omega^2}{c^2}\textbf{d}\cdot\mathrm{\widehat{\textbf{G}}}(\textbf{r}, \textbf{r}_\mathrm{QD}; \omega)(\omega + \omega_\mathrm{QD})/\varepsilon_0}{\omega_\mathrm{QD}^2 - \omega^2 - \frac{\omega^2}{c^2}\textbf{d}\cdot\widehat{\textbf{G}}(\textbf{r}, \textbf{r}_\mathrm{QD}; \omega)\cdot \textbf{d}/\hslash\varepsilon_0}\bigg|^2.
\end{equation}
Eq.~\ref{eq:ffspectrum} offers an intuitive picture since it incorporates the propagator $\mathrm{\widehat{\textbf{G}}}(\textbf{r}, \textbf{r}_\mathrm{QD}; \omega)$, that delineates the photon propagation to any point $\textbf{r} \left(\neq \textbf{r}_\mathrm{QD}\right)$ from the quantum emitter at $\textbf{r}_\mathrm{QD}$.

The far-field spectrum at Fig.~\ref{fig:fig_3}(c) and (d) provides interesting insight -- all the modes that appear in the near-field polarization spectrum in Fig.~\ref{fig:fig_3} (a) and (b), do not propagate to the far-field. The mode near topological transition (i.e., $\omega_{21}/\omega_\mathrm{t}\hspace{-4.5pt}\approx\hspace{-4.5pt}1.12$) is the only one that has far-field signature. Since, the modes at first anticrossing ($\omega/\omega_t\approx1.04$) are whispering gallery modes \cite{wu} that do not propagate to the far-field, and they show zero spectral signature of this mode at far-field. However, the mode at topological transition provides large Purcell factor while it transfers photon to the far-field. Unlike the traditional HMMs, this demonstrates large Purcell factor accompanied by radiation power enhancement. This is the central result of this work.

To further substantiate the aforementioned claim, harmonic decomposition of the Green's function in spherical harmonics has been performed (results not shown here). It is found that the first anticrossing has no dipole contribution (i.e., the contribution of $J=1$ is negligible), contrary to the mode at topological transition that has significant dipole contribution in the Green's function. We refer the mode at $\omega_{21}/\omega_\mathrm{t}\approx1.04$, is a pseudomode that appears as a contribution from higher order harmonics, similar to Ref.~\citep{vidal-prl}. 

%\section{Discussions and Conclusions}
To summarize, strong light-matter interaction in a hyperbolic cavity is demonstrated where two distinct anticrossings are present in the local polarization spectrum. The mechanisms responsible for the appearance of the two anticrossings are distinctive -- the first anti-crossing (i.e., the pseudo-mode) appears as contribution from higher order spherical harmonics of the cavity that is similar to the behaviour of plasmonic nano-particle, whilst the second anti-crossing appears due to the hyperbolicity of the hyperbolic cavity, at the topological transition of the cavity. The mode at topological transition propagates photon to the far-field as opposed to the pseudo-mode. This is quite unlike the case of quantum dots in the vicinity of planar layered hyperbolic metamaterials, where the large Purcell factors co-exist with the lowered radiation of the far field. The extension of the analysis for multiple quantum emitters is quite  straightforward via Dyson's equation~\cite{vidal-prl, gonzalez, kristensen} that involves the dressing of Green's function via incorporating the scattering by different quantum emitters and the rest of the system. 

%S\section{Acknowledgment}
We thank Andrey Bogdanov and Alexander Krasnok for discussion. The work was supported by the Ministry of Education and Science of the Russian Federation (Zadanie No. 3.1231.2014/K), the Russian Foundation for Basic Research, Grants of the President of Russian Federation, MK-5220.2015.2, MD- 7841.2015.2, and Dynasty foundation.
\bibliographystyle{apsrev4-1}
\bibliography{SCBIB}

\begin{thebibliography}{36}%
\makeatletter
\providecommand \@ifxundefined [1]{%
 \@ifx{#1\undefined}
}%
\providecommand \@ifnum [1]{%
 \ifnum #1\expandafter \@firstoftwo
 \else \expandafter \@secondoftwo
 \fi
}%
\providecommand \@ifx [1]{%
 \ifx #1\expandafter \@firstoftwo
 \else \expandafter \@secondoftwo
 \fi
}%
\providecommand \natexlab [1]{#1}%
\providecommand \enquote  [1]{``#1''}%
\providecommand \bibnamefont  [1]{#1}%
\providecommand \bibfnamefont [1]{#1}%
\providecommand \citenamefont [1]{#1}%
\providecommand \href@noop [0]{\@secondoftwo}%
\providecommand \href [0]{\begingroup \@sanitize@url \@href}%
\providecommand \@href[1]{\@@startlink{#1}\@@href}%
\providecommand \@@href[1]{\endgroup#1\@@endlink}%
\providecommand \@sanitize@url [0]{\catcode `\\12\catcode `\$12\catcode
  `\&12\catcode `\#12\catcode `\^12\catcode `\_12\catcode `\%12\relax}%
\providecommand \@@startlink[1]{}%
\providecommand \@@endlink[0]{}%
\providecommand \url  [0]{\begingroup\@sanitize@url \@url }%
\providecommand \@url [1]{\endgroup\@href {#1}{\urlprefix }}%
\providecommand \urlprefix  [0]{URL }%
\providecommand \Eprint [0]{\href }%
\providecommand \doibase [0]{http://dx.doi.org/}%
\providecommand \selectlanguage [0]{\@gobble}%
\providecommand \bibinfo  [0]{\@secondoftwo}%
\providecommand \bibfield  [0]{\@secondoftwo}%
\providecommand \translation [1]{[#1]}%
\providecommand \BibitemOpen [0]{}%
\providecommand \bibitemStop [0]{}%
\providecommand \bibitemNoStop [0]{.\EOS\space}%
\providecommand \EOS [0]{\spacefactor3000\relax}%
\providecommand \BibitemShut  [1]{\csname bibitem#1\endcsname}%
\let\auto@bib@innerbib\@empty
%</preamble>
\bibitem [{\citenamefont {Kavokin}\ \emph {et~al.}(2008)\citenamefont
  {Kavokin}, \citenamefont {Baumberg}, \citenamefont {Malpuech},\ and\
  \citenamefont {Laussy}}]{kavokin}%
  \BibitemOpen
  \bibfield  {author} {\bibinfo {author} {\bibfnamefont {A.~V.}\ \bibnamefont
  {Kavokin}}, \bibinfo {author} {\bibfnamefont {J.~J.}\ \bibnamefont
  {Baumberg}}, \bibinfo {author} {\bibfnamefont {G.}~\bibnamefont {Malpuech}},
  \ and\ \bibinfo {author} {\bibfnamefont {F.~P.}\ \bibnamefont {Laussy}},\
  }\href@noop {} {\emph {\bibinfo {title} {Microcavities}}}\ (\bibinfo
  {publisher} {Oxford University Press, USA},\ \bibinfo {year}
  {2008})\BibitemShut {NoStop}%
\bibitem [{\citenamefont {De~Liberato}\ \emph {et~al.}(2009)\citenamefont
  {De~Liberato}, \citenamefont {Gerace}, \citenamefont {Carusotto},\ and\
  \citenamefont {Ciuti}}]{deliberato}%
  \BibitemOpen
  \bibfield  {author} {\bibinfo {author} {\bibfnamefont {S.}~\bibnamefont
  {De~Liberato}}, \bibinfo {author} {\bibfnamefont {D.}~\bibnamefont {Gerace}},
  \bibinfo {author} {\bibfnamefont {I.}~\bibnamefont {Carusotto}}, \ and\
  \bibinfo {author} {\bibfnamefont {C.}~\bibnamefont {Ciuti}},\ }\href
  {\doibase 10.1103/PhysRevA.80.053810} {\bibfield  {journal} {\bibinfo
  {journal} {Phys. Rev. A}\ }\textbf {\bibinfo {volume} {80}},\ \bibinfo
  {pages} {053810} (\bibinfo {year} {2009})}\BibitemShut {NoStop}%
\bibitem [{\citenamefont {Beaudoin}\ \emph {et~al.}(2011)\citenamefont
  {Beaudoin}, \citenamefont {Gambetta},\ and\ \citenamefont
  {Blais}}]{beaudoin}%
  \BibitemOpen
  \bibfield  {author} {\bibinfo {author} {\bibfnamefont {F.}~\bibnamefont
  {Beaudoin}}, \bibinfo {author} {\bibfnamefont {J.~M.}\ \bibnamefont
  {Gambetta}}, \ and\ \bibinfo {author} {\bibfnamefont {A.}~\bibnamefont
  {Blais}},\ }\href {\doibase 10.1103/PhysRevA.84.043832} {\bibfield  {journal}
  {\bibinfo  {journal} {Phys. Rev. A}\ }\textbf {\bibinfo {volume} {84}},\
  \bibinfo {pages} {043832} (\bibinfo {year} {2011})}\BibitemShut {NoStop}%
\bibitem [{\citenamefont {Delga}\ \emph
  {et~al.}(2014{\natexlab{a}})\citenamefont {Delga}, \citenamefont {Feist},
  \citenamefont {Bravo-Abad},\ and\ \citenamefont {Garcia-Vidal}}]{vidal-prl}%
  \BibitemOpen
  \bibfield  {author} {\bibinfo {author} {\bibfnamefont {A.}~\bibnamefont
  {Delga}}, \bibinfo {author} {\bibfnamefont {J.}~\bibnamefont {Feist}},
  \bibinfo {author} {\bibfnamefont {J.}~\bibnamefont {Bravo-Abad}}, \ and\
  \bibinfo {author} {\bibfnamefont {F.~J.}\ \bibnamefont {Garcia-Vidal}},\
  }\href {\doibase 10.1103/PhysRevLett.112.253601} {\bibfield  {journal}
  {\bibinfo  {journal} {Phys. Rev. Lett.}\ }\textbf {\bibinfo {volume} {112}},\
  \bibinfo {pages} {253601} (\bibinfo {year} {2014}{\natexlab{a}})}\BibitemShut
  {NoStop}%
\bibitem [{\citenamefont {Van~Vlack}\ \emph {et~al.}(2012)\citenamefont
  {Van~Vlack}, \citenamefont {Kristensen},\ and\ \citenamefont
  {Hughes}}]{vvlac}%
  \BibitemOpen
  \bibfield  {author} {\bibinfo {author} {\bibfnamefont {C.}~\bibnamefont
  {Van~Vlack}}, \bibinfo {author} {\bibfnamefont {P.~T.}\ \bibnamefont
  {Kristensen}}, \ and\ \bibinfo {author} {\bibfnamefont {S.}~\bibnamefont
  {Hughes}},\ }\href {\doibase 10.1103/PhysRevB.85.075303} {\bibfield
  {journal} {\bibinfo  {journal} {Phys. Rev. B}\ }\textbf {\bibinfo {volume}
  {85}},\ \bibinfo {pages} {075303} (\bibinfo {year} {2012})}\BibitemShut
  {NoStop}%
\bibitem [{\citenamefont {Delga}\ \emph
  {et~al.}(2014{\natexlab{b}})\citenamefont {Delga}, \citenamefont {Feist},
  \citenamefont {Bravo-Abad},\ and\ \citenamefont {Garcia-Vidal}}]{delga}%
  \BibitemOpen
  \bibfield  {author} {\bibinfo {author} {\bibfnamefont {A.}~\bibnamefont
  {Delga}}, \bibinfo {author} {\bibfnamefont {J.}~\bibnamefont {Feist}},
  \bibinfo {author} {\bibfnamefont {J.}~\bibnamefont {Bravo-Abad}}, \ and\
  \bibinfo {author} {\bibfnamefont {F.~J.}\ \bibnamefont {Garcia-Vidal}},\
  }\href {http://stacks.iop.org/2040-8986/16/i=11/a=114018} {\bibfield
  {journal} {\bibinfo  {journal} {Journal of Optics}\ }\textbf {\bibinfo
  {volume} {16}},\ \bibinfo {pages} {114018} (\bibinfo {year}
  {2014}{\natexlab{b}})}\BibitemShut {NoStop}%
\bibitem [{\citenamefont {Wang}\ \emph {et~al.}(2014)\citenamefont {Wang},
  \citenamefont {Vasa}, \citenamefont {Sommer}, \citenamefont {Sio},
  \citenamefont {Gross}, \citenamefont {Vogelgesang},\ and\ \citenamefont
  {Lienau}}]{wang}%
  \BibitemOpen
  \bibfield  {author} {\bibinfo {author} {\bibfnamefont {W.}~\bibnamefont
  {Wang}}, \bibinfo {author} {\bibfnamefont {P.}~\bibnamefont {Vasa}}, \bibinfo
  {author} {\bibfnamefont {E.}~\bibnamefont {Sommer}}, \bibinfo {author}
  {\bibfnamefont {A.~D.}\ \bibnamefont {Sio}}, \bibinfo {author} {\bibfnamefont
  {P.}~\bibnamefont {Gross}}, \bibinfo {author} {\bibfnamefont
  {R.}~\bibnamefont {Vogelgesang}}, \ and\ \bibinfo {author} {\bibfnamefont
  {C.}~\bibnamefont {Lienau}},\ }\href
  {http://stacks.iop.org/2040-8986/16/i=11/a=114021} {\bibfield  {journal}
  {\bibinfo  {journal} {Journal of Optics}\ }\textbf {\bibinfo {volume} {16}},\
  \bibinfo {pages} {114021} (\bibinfo {year} {2014})}\BibitemShut {NoStop}%
\bibitem [{\citenamefont {Shalabney}\ \emph {et~al.}(2015)\citenamefont
  {Shalabney}, \citenamefont {George}, \citenamefont {Hutchison}, \citenamefont
  {Pupillo}, \citenamefont {Genet},\ and\ \citenamefont {Ebbesen}}]{shalabny}%
  \BibitemOpen
  \bibfield  {author} {\bibinfo {author} {\bibfnamefont {A.}~\bibnamefont
  {Shalabney}}, \bibinfo {author} {\bibfnamefont {J.}~\bibnamefont {George}},
  \bibinfo {author} {\bibfnamefont {J.}~\bibnamefont {Hutchison}}, \bibinfo
  {author} {\bibfnamefont {G.}~\bibnamefont {Pupillo}}, \bibinfo {author}
  {\bibfnamefont {C.}~\bibnamefont {Genet}}, \ and\ \bibinfo {author}
  {\bibfnamefont {T.~W.}\ \bibnamefont {Ebbesen}},\ }\href {\doibase
  10.1038/ncomms6981} {\bibfield  {journal} {\bibinfo  {journal} {Nature
  communications}\ }\textbf {\bibinfo {volume} {6}},\ \bibinfo {pages} {5981}
  (\bibinfo {year} {2015})}\BibitemShut {NoStop}%
\bibitem [{\citenamefont {Akahane}\ \emph {et~al.}(2003)\citenamefont
  {Akahane}, \citenamefont {Asano}, \citenamefont {Song},\ and\ \citenamefont
  {Noda.}}]{akahane}%
  \BibitemOpen
  \bibfield  {author} {\bibinfo {author} {\bibfnamefont {Y.}~\bibnamefont
  {Akahane}}, \bibinfo {author} {\bibfnamefont {T.}~\bibnamefont {Asano}},
  \bibinfo {author} {\bibfnamefont {B.-S.}\ \bibnamefont {Song}}, \ and\
  \bibinfo {author} {\bibfnamefont {S.}~\bibnamefont {Noda.}},\ }\href
  {\doibase 10.1038/nature02063} {\bibfield  {journal} {\bibinfo  {journal}
  {Nature}\ }\textbf {\bibinfo {volume} {425}},\ \bibinfo {pages} {6961}
  (\bibinfo {year} {2003})}\BibitemShut {NoStop}%
\bibitem [{\citenamefont {Reithmaier}\ \emph {et~al.}(2004)\citenamefont
  {Reithmaier}, \citenamefont {Sek}, \citenamefont {Loffler}, \citenamefont
  {Hofmann}, \citenamefont {S.}, \citenamefont {Reitzenstein}, \citenamefont
  {Keldysh}, \citenamefont {Kulakovskii}, \citenamefont {Reinecke},\ and\
  \citenamefont {Forchel}}]{reithmaier}%
  \BibitemOpen
  \bibfield  {author} {\bibinfo {author} {\bibfnamefont {J.~P.}\ \bibnamefont
  {Reithmaier}}, \bibinfo {author} {\bibfnamefont {G.}~\bibnamefont {Sek}},
  \bibinfo {author} {\bibfnamefont {A.}~\bibnamefont {Loffler}}, \bibinfo
  {author} {\bibfnamefont {C.}~\bibnamefont {Hofmann}}, \bibinfo {author}
  {\bibfnamefont {K.}~\bibnamefont {S.}}, \bibinfo {author} {\bibfnamefont
  {S.}~\bibnamefont {Reitzenstein}}, \bibinfo {author} {\bibfnamefont {L.~V.}\
  \bibnamefont {Keldysh}}, \bibinfo {author} {\bibfnamefont {V.~D.}\
  \bibnamefont {Kulakovskii}}, \bibinfo {author} {\bibfnamefont {T.~L.}\
  \bibnamefont {Reinecke}}, \ and\ \bibinfo {author} {\bibfnamefont
  {A.}~\bibnamefont {Forchel}},\ }\href {\doibase 10.1038/nature02969}
  {\bibfield  {journal} {\bibinfo  {journal} {Nature}\ }\textbf {\bibinfo
  {volume} {432}},\ \bibinfo {pages} {197} (\bibinfo {year}
  {2004})}\BibitemShut {NoStop}%
\bibitem [{\citenamefont {Torma}\ and\ \citenamefont {Barnes}(2015)}]{torma}%
  \BibitemOpen
  \bibfield  {author} {\bibinfo {author} {\bibfnamefont {P.}~\bibnamefont
  {Torma}}\ and\ \bibinfo {author} {\bibfnamefont {W.~L.}\ \bibnamefont
  {Barnes}},\ }\href {http://stacks.iop.org/0034-4885/78/i=1/a=013901}
  {\bibfield  {journal} {\bibinfo  {journal} {Reports on Progress in Physics}\
  }\textbf {\bibinfo {volume} {78}},\ \bibinfo {pages} {013901} (\bibinfo
  {year} {2015})}\BibitemShut {NoStop}%
\bibitem [{\citenamefont {Savasta}\ \emph {et~al.}(2010)\citenamefont
  {Savasta}, \citenamefont {Saija}, \citenamefont {Ridolfo}, \citenamefont
  {Di~Stefano}, \citenamefont {Denti},\ and\ \citenamefont
  {Borghese}}]{savasta}%
  \BibitemOpen
  \bibfield  {author} {\bibinfo {author} {\bibfnamefont {S.}~\bibnamefont
  {Savasta}}, \bibinfo {author} {\bibfnamefont {R.}~\bibnamefont {Saija}},
  \bibinfo {author} {\bibfnamefont {A.}~\bibnamefont {Ridolfo}}, \bibinfo
  {author} {\bibfnamefont {O.}~\bibnamefont {Di~Stefano}}, \bibinfo {author}
  {\bibfnamefont {P.}~\bibnamefont {Denti}}, \ and\ \bibinfo {author}
  {\bibfnamefont {F.}~\bibnamefont {Borghese}},\ }\href {\doibase
  10.1021/nn100585h} {\bibfield  {journal} {\bibinfo  {journal} {ACS Nano}\
  }\textbf {\bibinfo {volume} {4}},\ \bibinfo {pages} {6369} (\bibinfo {year}
  {2010})},\ \bibinfo {note} {pMID: 21028780}\BibitemShut {NoStop}%
\bibitem [{\citenamefont {Hakami}\ \emph {et~al.}(2014)\citenamefont {Hakami},
  \citenamefont {Wang},\ and\ \citenamefont {Zubairy}}]{zubairy}%
  \BibitemOpen
  \bibfield  {author} {\bibinfo {author} {\bibfnamefont {J.}~\bibnamefont
  {Hakami}}, \bibinfo {author} {\bibfnamefont {L.}~\bibnamefont {Wang}}, \ and\
  \bibinfo {author} {\bibfnamefont {M.~S.}\ \bibnamefont {Zubairy}},\ }\href
  {\doibase 10.1103/PhysRevA.89.053835} {\bibfield  {journal} {\bibinfo
  {journal} {Phys. Rev. A}\ }\textbf {\bibinfo {volume} {89}},\ \bibinfo
  {pages} {053835} (\bibinfo {year} {2014})}\BibitemShut {NoStop}%
\bibitem [{\citenamefont {Chantharasupawong}\ \emph {et~al.}(2014)\citenamefont
  {Chantharasupawong}, \citenamefont {Tetard},\ and\ \citenamefont
  {Thomas}}]{chantharasupawong}%
  \BibitemOpen
  \bibfield  {author} {\bibinfo {author} {\bibfnamefont {P.}~\bibnamefont
  {Chantharasupawong}}, \bibinfo {author} {\bibfnamefont {L.}~\bibnamefont
  {Tetard}}, \ and\ \bibinfo {author} {\bibfnamefont {J.}~\bibnamefont
  {Thomas}},\ }\href {\doibase 10.1021/jp506091k} {\bibfield  {journal}
  {\bibinfo  {journal} {The Jour. of Phys. Chem. C}\ }\textbf {\bibinfo
  {volume} {118}},\ \bibinfo {pages} {23954} (\bibinfo {year}
  {2014})}\BibitemShut {NoStop}%
\bibitem [{\citenamefont {Wu}\ \emph {et~al.}(2014{\natexlab{a}})\citenamefont
  {Wu}, \citenamefont {Cheng}, \citenamefont {Wu},\ and\ \citenamefont
  {Liu}}]{wu-oi}%
  \BibitemOpen
  \bibfield  {author} {\bibinfo {author} {\bibfnamefont {D.}~\bibnamefont
  {Wu}}, \bibinfo {author} {\bibfnamefont {Y.}~\bibnamefont {Cheng}}, \bibinfo
  {author} {\bibfnamefont {X.}~\bibnamefont {Wu}}, \ and\ \bibinfo {author}
  {\bibfnamefont {X.}~\bibnamefont {Liu}},\ }\href {\doibase
  10.1364/JOSAB.31.002273} {\bibfield  {journal} {\bibinfo  {journal} {J. Opt.
  Soc. Am. B}\ }\textbf {\bibinfo {volume} {31}},\ \bibinfo {pages} {2273}
  (\bibinfo {year} {2014}{\natexlab{a}})}\BibitemShut {NoStop}%
\bibitem [{\citenamefont {Cacciola}\ \emph {et~al.}(2014)\citenamefont
  {Cacciola}, \citenamefont {Di~Stefano}, \citenamefont {Stassi}, \citenamefont
  {Saija},\ and\ \citenamefont {Savasta}}]{cacciola}%
  \BibitemOpen
  \bibfield  {author} {\bibinfo {author} {\bibfnamefont {A.}~\bibnamefont
  {Cacciola}}, \bibinfo {author} {\bibfnamefont {O.}~\bibnamefont
  {Di~Stefano}}, \bibinfo {author} {\bibfnamefont {R.}~\bibnamefont {Stassi}},
  \bibinfo {author} {\bibfnamefont {R.}~\bibnamefont {Saija}}, \ and\ \bibinfo
  {author} {\bibfnamefont {S.}~\bibnamefont {Savasta}},\ }\href {\doibase
  10.1021/nn504652w} {\bibfield  {journal} {\bibinfo  {journal} {ACS Nano}\
  }\textbf {\bibinfo {volume} {8}},\ \bibinfo {pages} {11483} (\bibinfo {year}
  {2014})},\ \bibinfo {note} {pMID: 25337782}\BibitemShut {NoStop}%
\bibitem [{\citenamefont {Antosiewicz}\ \emph {et~al.}(2014)\citenamefont
  {Antosiewicz}, \citenamefont {Apell},\ and\ \citenamefont
  {Shegai}}]{antosiewicz}%
  \BibitemOpen
  \bibfield  {author} {\bibinfo {author} {\bibfnamefont {T.~J.}\ \bibnamefont
  {Antosiewicz}}, \bibinfo {author} {\bibfnamefont {S.~P.}\ \bibnamefont
  {Apell}}, \ and\ \bibinfo {author} {\bibfnamefont {T.}~\bibnamefont
  {Shegai}},\ }\href {\doibase 10.1021/ph500032d} {\bibfield  {journal}
  {\bibinfo  {journal} {ACS Photonics}\ }\textbf {\bibinfo {volume} {1}},\
  \bibinfo {pages} {454} (\bibinfo {year} {2014})}\BibitemShut {NoStop}%
\bibitem [{\citenamefont {Uemoto}\ and\ \citenamefont {Ajiki}(2014)}]{uemoto}%
  \BibitemOpen
  \bibfield  {author} {\bibinfo {author} {\bibfnamefont {M.}~\bibnamefont
  {Uemoto}}\ and\ \bibinfo {author} {\bibfnamefont {H.}~\bibnamefont {Ajiki}},\
  }\href {\doibase 10.1364/OE.22.022470} {\bibfield  {journal} {\bibinfo
  {journal} {Opt. Express}\ }\textbf {\bibinfo {volume} {22}},\ \bibinfo
  {pages} {22470} (\bibinfo {year} {2014})}\BibitemShut {NoStop}%
\bibitem [{\citenamefont {Thongrattanasiri}\ \emph {et~al.}(2013)\citenamefont
  {Thongrattanasiri}, \citenamefont {Manjavacas}, \citenamefont {Nordlander},\
  and\ \citenamefont {de~Abajo}}]{thongrattanasiri}%
  \BibitemOpen
  \bibfield  {author} {\bibinfo {author} {\bibfnamefont {S.}~\bibnamefont
  {Thongrattanasiri}}, \bibinfo {author} {\bibfnamefont {A.}~\bibnamefont
  {Manjavacas}}, \bibinfo {author} {\bibfnamefont {P.}~\bibnamefont
  {Nordlander}}, \ and\ \bibinfo {author} {\bibfnamefont {F.~J.~G.}\
  \bibnamefont {de~Abajo}},\ }\href {\doibase 10.1002/lpor.201200101}
  {\bibfield  {journal} {\bibinfo  {journal} {Laser and Photonics Reviews}\
  }\textbf {\bibinfo {volume} {7}},\ \bibinfo {pages} {297} (\bibinfo {year}
  {2013})}\BibitemShut {NoStop}%
\bibitem [{\citenamefont {Ghosh~Chaudhuri}\ and\ \citenamefont
  {Paria}(2011)}]{GhoshChaudhuri2011}%
  \BibitemOpen
  \bibfield  {author} {\bibinfo {author} {\bibfnamefont {R.}~\bibnamefont
  {Ghosh~Chaudhuri}}\ and\ \bibinfo {author} {\bibfnamefont {S.}~\bibnamefont
  {Paria}},\ }\href {http://pubs.acs.org/doi/pdfplus/10.1021/cr100449n}
  {\bibfield  {journal} {\bibinfo  {journal} {Chemical reviews}\ }\textbf
  {\bibinfo {volume} {112}},\ \bibinfo {pages} {2373} (\bibinfo {year}
  {2011})}\BibitemShut {NoStop}%
\bibitem [{\citenamefont {Chen}\ \emph {et~al.}(2012)\citenamefont {Chen},
  \citenamefont {Soric},\ and\ \citenamefont {Alu}}]{Chen2012}%
  \BibitemOpen
  \bibfield  {author} {\bibinfo {author} {\bibfnamefont {P.-Y.}\ \bibnamefont
  {Chen}}, \bibinfo {author} {\bibfnamefont {J.}~\bibnamefont {Soric}}, \ and\
  \bibinfo {author} {\bibfnamefont {A.}~\bibnamefont {Alu}},\ }\href
  {http://onlinelibrary.wiley.com/doi/10.1002/adma.201202624/abstract}
  {\bibfield  {journal} {\bibinfo  {journal} {Advanced Materials}\ }\textbf
  {\bibinfo {volume} {24}},\ \bibinfo {pages} {OP281} (\bibinfo {year}
  {2012})}\BibitemShut {NoStop}%
\bibitem [{\citenamefont {Rho}\ \emph {et~al.}(2010)\citenamefont {Rho},
  \citenamefont {Ye}, \citenamefont {Xiong}, \citenamefont {Yin}, \citenamefont
  {Liu}, \citenamefont {Choi}, \citenamefont {Bartal},\ and\ \citenamefont
  {Zhang}}]{Rho2010}%
  \BibitemOpen
  \bibfield  {author} {\bibinfo {author} {\bibfnamefont {J.}~\bibnamefont
  {Rho}}, \bibinfo {author} {\bibfnamefont {Z.}~\bibnamefont {Ye}}, \bibinfo
  {author} {\bibfnamefont {Y.}~\bibnamefont {Xiong}}, \bibinfo {author}
  {\bibfnamefont {X.}~\bibnamefont {Yin}}, \bibinfo {author} {\bibfnamefont
  {Z.}~\bibnamefont {Liu}}, \bibinfo {author} {\bibfnamefont {H.}~\bibnamefont
  {Choi}}, \bibinfo {author} {\bibfnamefont {G.}~\bibnamefont {Bartal}}, \ and\
  \bibinfo {author} {\bibfnamefont {X.}~\bibnamefont {Zhang}},\ }\href
  {http://dx.doi.org/10.1038/ncomms1148} {\bibfield  {journal} {\bibinfo
  {journal} {Nature communications}\ }\textbf {\bibinfo {volume} {1}},\
  \bibinfo {pages} {143} (\bibinfo {year} {2010})}\BibitemShut {NoStop}%
\bibitem [{\citenamefont {Cortes}\ \emph {et~al.}(2012)\citenamefont {Cortes},
  \citenamefont {Newman}, \citenamefont {Molesky},\ and\ \citenamefont
  {Jacob}}]{cortes}%
  \BibitemOpen
  \bibfield  {author} {\bibinfo {author} {\bibfnamefont {C.~L.}\ \bibnamefont
  {Cortes}}, \bibinfo {author} {\bibfnamefont {W.}~\bibnamefont {Newman}},
  \bibinfo {author} {\bibfnamefont {S.}~\bibnamefont {Molesky}}, \ and\
  \bibinfo {author} {\bibfnamefont {Z.}~\bibnamefont {Jacob}},\ }\href
  {http://stacks.iop.org/2040-8986/14/i=6/a=063001} {\bibfield  {journal}
  {\bibinfo  {journal} {Journal of Optics}\ }\textbf {\bibinfo {volume} {14}},\
  \bibinfo {pages} {063001} (\bibinfo {year} {2012})}\BibitemShut {NoStop}%
\bibitem [{\citenamefont {Jacob}\ \emph {et~al.}(2012)\citenamefont {Jacob},
  \citenamefont {Smolyaninov},\ and\ \citenamefont {Narimanov}}]{zubin}%
  \BibitemOpen
  \bibfield  {author} {\bibinfo {author} {\bibfnamefont {Z.}~\bibnamefont
  {Jacob}}, \bibinfo {author} {\bibfnamefont {I.~I.}\ \bibnamefont
  {Smolyaninov}}, \ and\ \bibinfo {author} {\bibfnamefont {E.~E.}\ \bibnamefont
  {Narimanov}},\ }\href {\doibase http://dx.doi.org/10.1063/1.4710548}
  {\bibfield  {journal} {\bibinfo  {journal} {Applied Physics Letters}\
  }\textbf {\bibinfo {volume} {100}},\ \bibinfo {eid} {181105} (\bibinfo {year}
  {2012})}\BibitemShut {NoStop}%
\bibitem [{\citenamefont {Poddubny}\ \emph
  {et~al.}(2013{\natexlab{a}})\citenamefont {Poddubny}, \citenamefont {Iorsh},
  \citenamefont {Belov},\ and\ \citenamefont {Kivshar}}]{vanya}%
  \BibitemOpen
  \bibfield  {author} {\bibinfo {author} {\bibfnamefont {A.}~\bibnamefont
  {Poddubny}}, \bibinfo {author} {\bibfnamefont {I.}~\bibnamefont {Iorsh}},
  \bibinfo {author} {\bibfnamefont {P.}~\bibnamefont {Belov}}, \ and\ \bibinfo
  {author} {\bibfnamefont {Y.}~\bibnamefont {Kivshar}},\ }\href {\doibase
  10.1038/nphoton.2013.243} {\bibfield  {journal} {\bibinfo  {journal} {Nat
  Photon}\ }\textbf {\bibinfo {volume} {7}},\ \bibinfo {pages} {948} (\bibinfo
  {year} {2013}{\natexlab{a}})}\BibitemShut {NoStop}%
\bibitem [{\citenamefont {Poddubny}\ \emph {et~al.}(2011)\citenamefont
  {Poddubny}, \citenamefont {Belov},\ and\ \citenamefont
  {Kivshar}}]{puddubny_pra}%
  \BibitemOpen
  \bibfield  {author} {\bibinfo {author} {\bibfnamefont {A.~N.}\ \bibnamefont
  {Poddubny}}, \bibinfo {author} {\bibfnamefont {P.~A.}\ \bibnamefont {Belov}},
  \ and\ \bibinfo {author} {\bibfnamefont {Y.~S.}\ \bibnamefont {Kivshar}},\
  }\href {\doibase 10.1103/PhysRevA.84.023807} {\bibfield  {journal} {\bibinfo
  {journal} {Phys. Rev. A}\ }\textbf {\bibinfo {volume} {84}},\ \bibinfo
  {pages} {023807} (\bibinfo {year} {2011})}\BibitemShut {NoStop}%
\bibitem [{\citenamefont {Poddubny}\ \emph
  {et~al.}(2013{\natexlab{b}})\citenamefont {Poddubny}, \citenamefont {Belov},\
  and\ \citenamefont {Kivshar}}]{puddubny_springer}%
  \BibitemOpen
  \bibfield  {author} {\bibinfo {author} {\bibfnamefont {A.~N.}\ \bibnamefont
  {Poddubny}}, \bibinfo {author} {\bibfnamefont {P.~A.}\ \bibnamefont {Belov}},
  \ and\ \bibinfo {author} {\bibfnamefont {Y.~S.}\ \bibnamefont {Kivshar}},\
  }\href {\doibase 10.1117/12.2023857} {\bibfield  {journal} {\bibinfo
  {journal} {Proc. SPIE}\ }\textbf {\bibinfo {volume} {8806}},\ \bibinfo
  {pages} {88060T} (\bibinfo {year} {2013}{\natexlab{b}})}\BibitemShut
  {NoStop}%
\bibitem [{\citenamefont {Krishnamoorthy}\ \emph {et~al.}(2012)\citenamefont
  {Krishnamoorthy}, \citenamefont {Jacob}, \citenamefont {Narimanov},
  \citenamefont {Kretzschmar},\ and\ \citenamefont {Menon}}]{krishnamoorthy}%
  \BibitemOpen
  \bibfield  {author} {\bibinfo {author} {\bibfnamefont {H.~N.~S.}\
  \bibnamefont {Krishnamoorthy}}, \bibinfo {author} {\bibfnamefont
  {Z.}~\bibnamefont {Jacob}}, \bibinfo {author} {\bibfnamefont
  {E.}~\bibnamefont {Narimanov}}, \bibinfo {author} {\bibfnamefont
  {I.}~\bibnamefont {Kretzschmar}}, \ and\ \bibinfo {author} {\bibfnamefont
  {V.~M.}\ \bibnamefont {Menon}},\ }\href {\doibase 10.1126/science.1219171}
  {\bibfield  {journal} {\bibinfo  {journal} {Science}\ }\textbf {\bibinfo
  {volume} {336}},\ \bibinfo {pages} {205} (\bibinfo {year}
  {2012})}\BibitemShut {NoStop}%
\bibitem [{\citenamefont {Li}\ \emph {et~al.}(1994)\citenamefont {Li},
  \citenamefont {Kooi}, \citenamefont {Leong},\ and\ \citenamefont {Yee}}]{Li}%
  \BibitemOpen
  \bibfield  {author} {\bibinfo {author} {\bibfnamefont {L.~W.}\ \bibnamefont
  {Li}}, \bibinfo {author} {\bibfnamefont {P.~S.}\ \bibnamefont {Kooi}},
  \bibinfo {author} {\bibfnamefont {M.~S.}\ \bibnamefont {Leong}}, \ and\
  \bibinfo {author} {\bibfnamefont {T.~S.}\ \bibnamefont {Yee}},\ }\href
  {\doibase 10.1109/22.339756} {\bibfield  {journal} {\bibinfo  {journal} {IEEE
  Tran. on Mic. Theo. and Tec.}\ }\textbf {\bibinfo {volume} {42}},\ \bibinfo
  {pages} {2302} (\bibinfo {year} {1994})}\BibitemShut {NoStop}%
\bibitem [{\citenamefont {Gu}\ \emph {et~al.}(2014)\citenamefont {Gu},
  \citenamefont {Tumkur}, \citenamefont {Zhu},\ and\ \citenamefont
  {Noginov}}]{noginov}%
  \BibitemOpen
  \bibfield  {author} {\bibinfo {author} {\bibfnamefont {L.}~\bibnamefont
  {Gu}}, \bibinfo {author} {\bibfnamefont {T.~U.}\ \bibnamefont {Tumkur}},
  \bibinfo {author} {\bibfnamefont {G.}~\bibnamefont {Zhu}}, \ and\ \bibinfo
  {author} {\bibfnamefont {M.~A.}\ \bibnamefont {Noginov}},\ }\href {\doibase
  10.1038/srep04969} {\bibfield  {journal} {\bibinfo  {journal} {Scientific
  Reports}\ }\textbf {\bibinfo {volume} {4}},\ \bibinfo {pages} {4969}
  (\bibinfo {year} {2014})}\BibitemShut {NoStop}%
\bibitem [{\citenamefont {Novotny}\ and\ \citenamefont
  {Hecht}(2012)}]{novotny}%
  \BibitemOpen
  \bibfield  {author} {\bibinfo {author} {\bibfnamefont {L.}~\bibnamefont
  {Novotny}}\ and\ \bibinfo {author} {\bibfnamefont {B.}~\bibnamefont
  {Hecht}},\ }\href@noop {} {\emph {\bibinfo {title} {Principles of
  Nano-Optics}}}\ (\bibinfo  {publisher} {Cambridge University Press, USA, 2nd
  Ed.},\ \bibinfo {year} {2012})\BibitemShut {NoStop}%
\bibitem [{\citenamefont {Bethe}(1947)}]{bethe}%
  \BibitemOpen
  \bibfield  {author} {\bibinfo {author} {\bibfnamefont {H.~A.}\ \bibnamefont
  {Bethe}},\ }\href {\doibase 10.1103/PhysRev.72.339} {\bibfield  {journal}
  {\bibinfo  {journal} {Phys. Rev.}\ }\textbf {\bibinfo {volume} {72}},\
  \bibinfo {pages} {339} (\bibinfo {year} {1947})}\BibitemShut {NoStop}%
\bibitem [{\citenamefont {Dung}\ \emph {et~al.}(2003)\citenamefont {Dung},
  \citenamefont {Buhmann}, \citenamefont {Kn\"oll}, \citenamefont {Welsch},
  \citenamefont {Scheel},\ and\ \citenamefont {K\"astel}}]{welsh}%
  \BibitemOpen
  \bibfield  {author} {\bibinfo {author} {\bibfnamefont {H.~T.}\ \bibnamefont
  {Dung}}, \bibinfo {author} {\bibfnamefont {S.~Y.}\ \bibnamefont {Buhmann}},
  \bibinfo {author} {\bibfnamefont {L.}~\bibnamefont {Kn\"oll}}, \bibinfo
  {author} {\bibfnamefont {D.-G.}\ \bibnamefont {Welsch}}, \bibinfo {author}
  {\bibfnamefont {S.}~\bibnamefont {Scheel}}, \ and\ \bibinfo {author}
  {\bibfnamefont {J.}~\bibnamefont {K\"astel}},\ }\href {\doibase
  10.1103/PhysRevA.68.043816} {\bibfield  {journal} {\bibinfo  {journal} {Phys.
  Rev. A}\ }\textbf {\bibinfo {volume} {68}},\ \bibinfo {pages} {043816}
  (\bibinfo {year} {2003})}\BibitemShut {NoStop}%
\bibitem [{\citenamefont {Wu}\ \emph {et~al.}(2014{\natexlab{b}})\citenamefont
  {Wu}, \citenamefont {Salandrino}, \citenamefont {Ni},\ and\ \citenamefont
  {Zhang}}]{wu}%
  \BibitemOpen
  \bibfield  {author} {\bibinfo {author} {\bibfnamefont {C.}~\bibnamefont
  {Wu}}, \bibinfo {author} {\bibfnamefont {A.}~\bibnamefont {Salandrino}},
  \bibinfo {author} {\bibfnamefont {X.}~\bibnamefont {Ni}}, \ and\ \bibinfo
  {author} {\bibfnamefont {X.}~\bibnamefont {Zhang}},\ }\href {\doibase
  10.1103/PhysRevX.4.021015} {\bibfield  {journal} {\bibinfo  {journal} {Phys.
  Rev. X}\ }\textbf {\bibinfo {volume} {4}},\ \bibinfo {pages} {021015}
  (\bibinfo {year} {2014}{\natexlab{b}})}\BibitemShut {NoStop}%
\bibitem [{\citenamefont {Gonzalez-Ballestero}\ \emph
  {et~al.}(2015)\citenamefont {Gonzalez-Ballestero}, \citenamefont {Feist},
  \citenamefont {Moreno},\ and\ \citenamefont {Garcia-Vidal}}]{gonzalez}%
  \BibitemOpen
  \bibfield  {author} {\bibinfo {author} {\bibfnamefont {C.}~\bibnamefont
  {Gonzalez-Ballestero}}, \bibinfo {author} {\bibfnamefont {J.}~\bibnamefont
  {Feist}}, \bibinfo {author} {\bibfnamefont {E.}~\bibnamefont {Moreno}}, \
  and\ \bibinfo {author} {\bibfnamefont {F.~J.}\ \bibnamefont {Garcia-Vidal}},\
  }\href@noop {} {\  (\bibinfo {year} {2015})},\ \Eprint
  {http://arxiv.org/abs/1502.04905} {arXiv:1502.04905 [cond-mat.mes-hall]}
  \BibitemShut {NoStop}%
\bibitem [{\citenamefont {Kristensen}\ \emph {et~al.}(2011)\citenamefont
  {Kristensen}, \citenamefont {M\o{}rk}, \citenamefont {Lodahl},\ and\
  \citenamefont {Hughes}}]{kristensen}%
  \BibitemOpen
  \bibfield  {author} {\bibinfo {author} {\bibfnamefont {P.~T.}\ \bibnamefont
  {Kristensen}}, \bibinfo {author} {\bibfnamefont {J.}~\bibnamefont {M\o{}rk}},
  \bibinfo {author} {\bibfnamefont {P.}~\bibnamefont {Lodahl}}, \ and\ \bibinfo
  {author} {\bibfnamefont {S.}~\bibnamefont {Hughes}},\ }\href {\doibase
  10.1103/PhysRevB.83.075305} {\bibfield  {journal} {\bibinfo  {journal} {Phys.
  Rev. B}\ }\textbf {\bibinfo {volume} {83}},\ \bibinfo {pages} {075305}
  (\bibinfo {year} {2011})}\BibitemShut {NoStop}%
\end{thebibliography}%
\end{document}